# Novel Rubidium Poly-Nitrogen Materials at High Pressure

Ashley S. Williams,[1] Brad A. Steele,[1] and Ivan I. Oleynik[1, a)]
*Department of Physics, University of South Florida, Tampa, FL 33620*

First-principles crystal structure search is performed to predict novel rubidium poly-nitrogen materials at high pressure by varying the stoichiometry, i. e. relative quantities of the constituent rubidium and nitrogen atoms. Three compounds of high nitrogen content, $RbN_5$, $RbN_2$, and $Rb_4N_6$, are discovered. Rubidium pentazolate ($RbN_5$) becomes thermodynamically stable at pressures above 30 GPa. The charge transfer from Rb to N atoms enables aromaticity in cyclo-$N_5^-$ while increasing the ionic bonding in the crystal. Rubidium pentazolate can be synthesized by compressing rubidium azide ($RbN_3$) and nitrogen ($N_2$) precursors above 9.42 GPa, and its experimental discovery is aided by calculating the Raman spectrum and identifying the features attributed to $N_5^-$ modes. The two other interesting compounds, $RbN_2$ containing infinitely-long single-bonded nitrogen chains, and $Rb_4N_6$ consisting of single-bonded $N_6$ hexazine rings, become thermodynamically stable at pressures exceeding 60 GPa. In addition to the compounds with high nitrogen content, $Rb_3N_3$, a new compound with 1:1 RbN stoichiometry containing bent $N_3$ azides is found to exist at high pressures.

## I. INTRODUCTION

Poly-nitrogen materials are being actively investigated[1–6] with the goal to obtain stable high-nitrogen content compounds, which are expected to outperform conventional all-organic energetic materials (EMs). The pure single-bonded cubic gauche (cg) polynitrogen crystal[7] would be the leading candidate, as the decomposition of the single-bonded condensed-phase nitrogen to the gas-phase triple bonded diatomic $N_2$ molecules results in enormous release of energy. In 2004, Eremets *et al.*[8] successfully synthesized the single-bonded cg phase of polymeric nitrogen, that was originally predicted to be energetically favorable over the epsilon phase of $N_2$ above 50 GPa[9]. In 2014, Tomasino *et al.*[10] discovered a new layered polymeric phase of nitrogen. Unfortunately, the synthesis of polymeric nitrogen requires excessive heating above 2000 K and enormous pressures above 100 GPa. Moreover, its recovery at ambient conditions is challenging[11]. For this reason, new approaches for the synthesis of high nitrogen content materials are urgently sought[1–3,6,12–14].

Introduction of other elements into polynitrogen-containing crystals appears to be a viable route to promote metastability while reducing synthesis pressure[12,13,15]. Previously, alkali polynitrogen materials including sodium[13], cesium[12,16], and lithium[17] polynitrides have been investigated at high pressures. In particular, cesium pentazolate $CsN_5$, consisting of pentazolate $N_5^-$ anions and $Cs^+$ cations was predicted, and then synthesized[12]. The presence of alkali metal atoms facilitates the electron transfer to nitrogen $N_5$ clusters, thus making cyclo-$N_5^-$ aromatic. Previously studied alkali nitrides $XN_y$, X = {Cs, Na, Li, K} display similarities: some of them contain $N_5$ pentazolates at a specific stoichiometry $XN_5$ or infinite nitrogen chains in $XN_2$ stoichiometry[12,13,16–18]. First-principles calculations show that cesium pentazolate[12,16] and lithium pentazolate[17] are stable above 15 GPa whereas sodium pentazolate is predicted to be stable above 20 GPa. The cesium pentazolate material was synthesized in a diamond anvil cell by compressing and laser heating $CsN_3$ and $N_2$ to pressures near 60 GPa[12]. High-nitrogen compound $X_4N_6$ containing $N_6$ hexazine rings is predicted to exist in the case of cesium and potassium, whereas no such structures are found in case of $Li_xN_y$ and $Na_xN_y$ materials[12,17,18].

Based on the similarities and differences between alkali polynitrides, the interesting question is whether rubidium pentazolate exists and whether it could be synthesized under similar conditions as $CsN_5$ using possible precursor mixture, $RbN_3+N_2$. The barrier to decomposition of cyclo-$N_5^-$ into $N_3^- + N_2$ is significant, 92 kJ/mol. Therefore it is expected to be metastable at ambient conditions making it a promising candidate for high nitrogen content EMs.[19,20] The relative stability of cyclo-$N_5^-$ allowed the synthesis of cyclo-$N_5^-$ compounds at ambient conditions.[21–24]

In this work, rubidium polynitrogens are investigated with the goal to explore novel polynitrogen EMs. Being similar to cesium, rubidium polynitrogens are interesting as for both elements, *d* electrons of the core begin to play a prominent role at high pressures. An indirect indication of the expected diversity in rubidium polynitrogen compounds is the complex phase diagram of pure Rb, which possesses six phases between 0 and 48 GPa[25,26]. This is in contrast to Na that has only two stable phases between 0 GPa and 100 GPa with the phase transition from bcc to fcc occurring at sufficiently high pressure 60 GPa[27]. It is suggested that the presence of occupied *3d* orbital results in destabilization of the bcc and fcc structures at high pressures[25]. The goal of this work is to predict and characterize new rubidium polynitrogen materials, including but not limited to $RbN_5$ and $RbN_2$, as well as to discuss future experimental realization of our predictions by suggesting possible precursors and corresponding pressures to initiate the high pressure chemistry.

---

a) Electronic mail: oleynik@usf.edu

## II. COMPUTATIONAL METHODS

A search for new compounds of varying $Rb_xN_y$ stoichiometry is performed using the first-principles evolutionary structure prediction method USPEX[28–30] at four pressures, 0, 30, 60 and 100 GPa. For each pressure the search begins by generating crystal structures containing variable amounts of elemental rubidium and nitrogen with 8 - 16 atoms in the unit cell. The crystals in the first generation are built with a random composition and lattice parameters. Density functional theory (DFT) is then used to minimize the total energy at fixed hydrostatic pressure by relaxing the atomic coordinates and the unit cell parameters of each structure of the entire population. The structures are then ranked by their formation enthalpy, those with the lowest formation enthalpy being the most favorable. A new generation of crystals is then produced with a combination of the lowest enthalpy structures from the previous generation, new random structures, and by applying variation operators, such as atom mutations and lattice permutations, to the lowest enthalpy structures to make a set of new crystals. This cyclic process of creating and optimizing new generations repeats until the best structures do not change for the last ten generations.

If applied properly, the USPEX method is very robust. In particular, we search for 99 compositions of $Rb_xN_y$ over a wide range of pressure. Taking into account such an exhaustive set of stoichiometries and the relative simplicity of the chemistry of the binary compounds, it is very likely the method found all the important compounds consisting of a reasonable number of atoms per unit cell. In two cases, a molecular search was also conducted in addition to the standard USPEX variable composition search to determine if larger crystals would lower the enthalpy compared to the smaller molecular crystals that were found during the variable composition search.

The convex hull, a formation enthalpy versus composition plot, is generated to determine which compounds are thermodynamically stable. The convex hull is constructed with the lowest formation enthalpy phases of each compound. If the compound lies on the convex hull, it is considered thermodynamically stable[31] and will not decompose into the constituent pure elements or other $Rb_xN_y$ compounds. The formation enthalpy is defined as the difference in enthalpy of a given compound and the sum of the enthalpies of the pure referenced structures, Rb and N, in their aggregate state at a given pressure and temperature. The reference structures of elemental N and Rb used to calculate the enthalpy of formation are α-$N_2$ and bcc Rb at 0 GPa, ε-$N_2$ and $I4/amd$ Rb at 30 GPa, and cg-N and $Cmca$ Rb at 60 GPa and 100 GPa. As the reference structures are not the focus of this paper and their phase diagrams are well known, pure Rb and N were not included in this USPEX search. The correct phases of rubidium have been predicted at high pressures using the USPEX algorithm[26] while a similar method was used by C.J. Pickard and R. J. Needs[32] to discover both experimentally known phases of nitrogen as well as new phases.

First-principles calculations are performed using DFT code VASP[33] and employing Perdew-Burke-Ernzerhof (PBE) generalized gradient approximation (GGA) to DFT[34], projector augmented wave (PAW) pseudopotentials, plane wave basis set with 350 eV energy cutoff, and k-point sampling with density 0.07 Å$^{-1}$. For each pressure, the lowest enthalpy structures and those close to the convex hull are then calculated at higher accuracy to compute the true convex hulls. As triply-bonded nitrogen has extremely small bond lengths, the hard nitrogen PAW pseudopotential with the core radius 0.582 Å is used. Correspondingly, the energy cutoff for the high accuracy calculations is raised to 1000 eV, while the k-point density is increased to 0.04 Å$^{-1}$. The chemical bonding (atomic charges and the bond orders) in novel rubidium polynitrogen compounds is analyzed using local combination of atomic orbitals (LCAO) code DMol[35,36]. The off-resonant Raman frequencies are obtained within frozen phonon approximation by calculating phonons at the gamma point, and their intensities – by calculating the derivatives of macroscopic dielectric polarizability tensor along the normal mode eigenvectors[37,38].

## III. RESULTS AND DISCUSSION

Our first-principles structure search of $Rb_xN_y$ compounds of varying stoichiometry produced several new rubidium polynitrogen structures at various pressures. The thermodynamically stable structures discovered in the search are displayed in the convex hulls (Fig. 1a) which are then used to produce the corresponding phase diagram (Fig. 1b). The crystal structures of several newly predicted $Rb_xN_y$ compounds are shown in Fig. 2.

At 0 GPa, there are two stable compounds: $RbN_2$-*P-1* consisting of $N_2$ molecules (Fig. 2a) and Rb atoms, as well as experimentally known rubidium azide ($RbN_3$) polymorph with *I4/mcm* symmetry (Fig. 2b)[39]. These two compounds, $RbN_2$ and $RbN_3$, remain stable at 15 GPa, $RbN_3$ transforming to the *C2/m* phase and eventually becoming metastable at pressures above 30 GPa. This study correctly predicts the $RbN_3$-*I4/mcm* phase to be the most energetically favorable at ambient conditions[39], as well as the pressure-induced phase transition of this phase to that of *C2/m* symmetry[40] in agreement with experiment. The validation of the USPEX method in the scope of this work is demonstrated by the correct prediction the $RbN_3$ phases listed above.

Wang *et al.*[41] found that $RbN_3$ undergoes the following phase transitions: to a *P-1* phase containing infinite chains of nitrogen at 30 GPa, and to *P6/mmm* phase containing $N_6$ hexazine rings at 50 GPa. This is not what we found in this work. At 30 GPa, the lowest enthalpy structure is still the *C2/m* phase. The *P6/mmm* phase is not on the convex hull at 60 GPa, see corresponding open violet circle in Figure 1a). Eventually, upon increasing pressure to 100 GPa, the $P6/mmm$ phase of $RbN_3$ with 6-membered hexazine rings becomes stable.

Several other structures appear above 15 GPa. One of them, $Rb_3N_3$-$P$-$1$ crystal with 1:1 stoichiometry, consisting of bent $N_3$ chains and three Rb atoms per each $N_3$ molecule (Fig. 2c)), remains the lowest formation enthalpy compound at pressures up to 38.4 GPa where it transforms to $Rb_4N_4$-$P$-$1$ crystal with the same 1:1 stoichiometry containing one-dimensional four-atom nitrogen chains ($N_4$) and isolated Rb atoms, see Fig. 2d).

A phase diagram for the phases with 1:1 stoichiometry is constructed to predict the interval of pressures where the novel $Rb_3N_3$-$P$-$1$ compound is stable. The three lowest RbN phases found during the USPEX search, $Rb_3N_3$-$P$-$1$, $Rb_4N_4$-$P$-$1$, $Rb_2N_2$-$P$-$1$-$a$, as well as another high-pressure polymorph $Rb_2N_2$-$P$-$1$-$b$, were optimized at fixed pressures in interval from 0 to 100 GPa. The $Rb_2N_2$ $P$-$1$-b crystal has the same symmetry $P$-$1$, as $Rb_2N_2$-$P$-$1$-$a$, but has different positions of the nitrogen atoms. The reference structure for the compression analysis is taken to be the $Rb_3N_3$-$P$-$1$ phase. As seen from Fig. 3, the $Rb_2N_2$-$P$-$1$-a phase, containing $N_2$ molecules and Rb atoms, is the lowest enthalpy phase only until 1.3 GPa. At 0 GPa, $Rb_2N_2$-$P$-$1$-a is 90 meV above the convex hull (1b)). From 1.3 GPa to 38.4 GPa the $Rb_3N_3$-$P$-$1$ phase with the bent $N_3$ molecules, see Fig. 2c), is the lowest enthalpy phase. At pressures above 38.4 GPa the $Rb_4N_4$-$P$-$1$ phase, consisting of bent $N_4$ molecules, is the lowest enthalpy phase, Fig. 2d).

On the rubidium-rich side of the convex hull, $Rb_3N_2$ - $C2/c$ and $Rb_3N$-$Pnma$ crystals are predicted to be stable at pressures above 15 GPa. They are composed of $N_2$ molecules and isolated Rb atoms. On the nitrogen-rich side, $RbN_5$-$Pc$ also appears on the convex hull at 15 GPa, which contains cyclo-$N_5^-$ pentazole- five-member rings of nitrogen atoms (Fig. 2f)). This phase of $RbN_5$ remains the lowest stable phase at pressures up to 60 GPa, above this pressure it transforms to $RbN_5$-$Pnma$.

At 60 GPa there is a phase transition of the $RbN_2$ crystal from the $P$-$1$ phase to the $Pbcn$ phase. This new phase, $RbN_2$-$Pbcn$, consists of infinite chains of nitrogen atoms (Fig. 2g). Another compound predicted in this study, $Rb_4N_6$-$P$-$1$ (Fig. 2h)), is close to the convex hull at 30 GPa, then appears on the convex hull at 60 GPa and remains stable to at least 100 GPa. This crystal, consisting of six-membered hexazine rings and having 4:6 stoichiometry, has analog in the family of cesium and potassium poly-nitrides: $Cs_4N_6$ becomes thermodynamically stable at pressures greater than 30 GPa and $K_4N_6$ becomes thermodynamically stable at pressures greater than 60 GPa.[12,18], but such stoichiometry has not been seen in the sodium[13] or lithium studies[17]. However, in the latter case, as well as in case of Cs, the compound of different stoichiometry has been found to contain hexazine $N_6$ rings: $LiN_3$ above 60 GPa[17], and $CsN_3$ from 80 GPa to at least 100 GPa[16].

The prediction of $RbN_5$ adds this pentazolate compound to the list of other alkali pentazolates: $LiN_5$[17], $NaN_5$[13], $CsN_5$[12,16], and $KN_5$[18]. Therefore, an important conclusion can be made: $XN_5$ pentazolate crystals are the general feature of nitrogen-alkali metal chemistry at high pressures. The $RbN_5$-$Pc$ phase becomes stable at relatively low pressure of 15 GPa, which is similar to the case of three other alkali-pentazolate compounds: $CsN_5$- stable at 15 GPa[12,17], $LiN_5$ - at 15 GPa[17], $KN_5$ - stable at 15 GPa[18] and $NaN_5$ - at 20 GPa[13]. In addition to pentazolates, infinite nitrogen chains are found in high pressure phase of $RbN_2$-$Pbcn$, which was also seen in case of sodium, cesium, and lithium polynitrides. The infinite chain structure of $RbN_2$ in the $Pbcn$ phase exists at pressures greater than 60 GPa.

To understand the structure and bonding in the novel $Rb_xN_y$ compounds, their charges and bond orders are calculated at 0 GPa and shown in Fig. 4. The N$_2$ molecules in RbN$_2$-*P-1* crystal carry a net charge of $-0.248e$ and have the N-N bond order of 2.05, which indicates a double bond. At ambient conditions the bond order for the azide in rubidium azide (RbN$_3$) is 1.97 while the charge on both outer N atoms of the azide is -0.508$e$ and the charge on the center atom is +0.397$e$. The average bond order of the N-N bond in bent N$_3$ azides in Rb$_3$N$_3$-*P-1* crystal (which is stable above 60 GPa) is 1.32, see Fig. 3. The smaller bond order compared to the double N-N bonds (bond order 2) of N$_3$ azide in RbN$_3$-*I4/mcm* is due to the filling of the anti-bonding orbitals by extra negative charge ( the charge on the azide N$_3$ chain in RbN$_3$-*I4/mcm* is -0.619 compared to -1.305 in Rb$_3$N$_3$-*P-1*). The lowest enthalpy phase of RbN between 60 and 100 GPa, Rb$_4$N$_4$-*P-1*, consists of bent N$_4$ molecules. The bond order of the N-N bonds in Rb$_4$N$_4$-*P-1* crystal involving the outer N atom, carrying -0.621$e$, and the inner nitrogen atom, carrying the charge -0.253$e$, is 1.17 and the bond order of the bond between the two inner nitrogen atoms is 1.04. The charges on the nitrogen atoms in the cyclo-N$_5$ ring of the RbN$_5$-*Pc* compound are -0.171$e$, -0.121$e$, -0.182$e$, -0.167$e$, and -0.156$e$ and the corresponding N-N bond orders range from 1.39 to 1.43, which demonstrates the aromatic nature of the pentazolate N$_5^-$ anion.

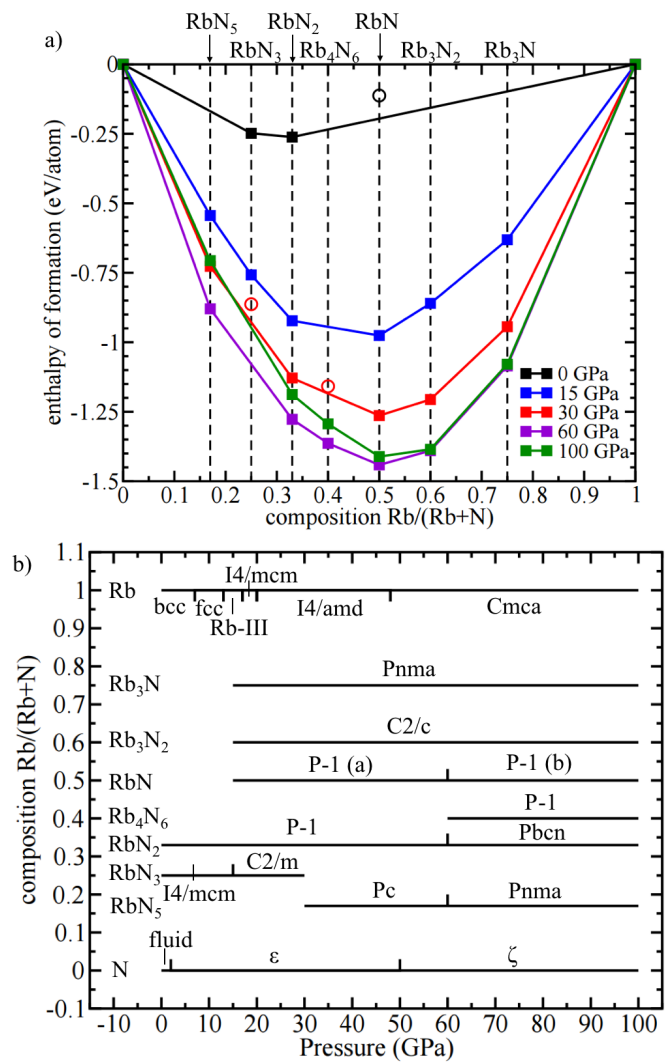

FIG. 1. a) Convex hulls at pressures 0, 15, 30, 60, and 100 GPa. Thermodynamically stable structures are marked by filled circles; metastable structures are marked with open circles. b) Composition versus pressure phase diagram of $Rb_xN_y$ crystals.

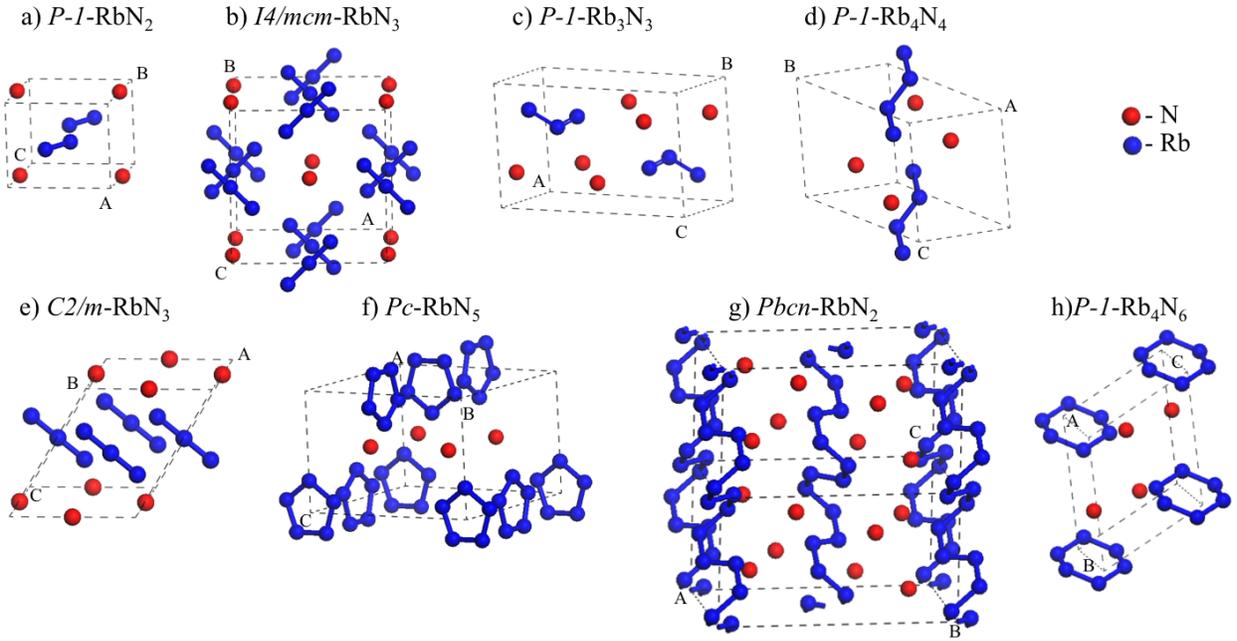

FIG. 2. Crystal structures of select rubidium polynitrides: a) RbN$_2$-*P-1* (stable at 0-30 GPa); b) RbN$_3$-*I4/mcm* (stable at 0-15 GPa); c) RbN-*P-1* (stable at 30-60 GPa); d) RbN-*P-1* (stable above 60 GPa); e) RbN$_3$-*C2/m* (stable at 15-30 GPa); f) RbN$_5$-*Pc* (stable at 30-60 GPa); g) RbN$_2$-*Pbcn* (stable above 60 GPa); h) Rb$_4$N$_6$-*P-1* (stable above 60 GPa)

The infinite N-N chains in RbN$_2$-*Pbcn* crystal (the lowest enthalpy phase above 60 GPa), have varying N-N bond orders from 0.99 to 1.35. The charges on N atoms follow a cyclic pattern with two N atoms having charge of -0.156$e$ while the next two nitrogen atoms – -0.335$e$ followed by two more nitrogen atoms of charge -0.156$e$ and so on. The irregular geometry of the nitrogen chains and the proximity of the nitrogen atoms to the rubidium atoms results in such irregular charge distribution.

The Rb$_4$N$_6$-*P-1* phase, which is thermodynamically stable at pressures exceeding 60 GPa, contains N$_6$ hexazine rings. These rings are made of symmetric groups of three nitrogen atoms with bond orders of 1.03, 1.05, and 1.02, with nearly identical bond lengths of 1.41 Å between $N_1$-$N_2$ and 1.4 Å between $N_2$-$N_3$ and $N_3$-$N_1$. The charges on N atoms in $N_6$ ring vary from -0.344$e$ to -0.371$e$.

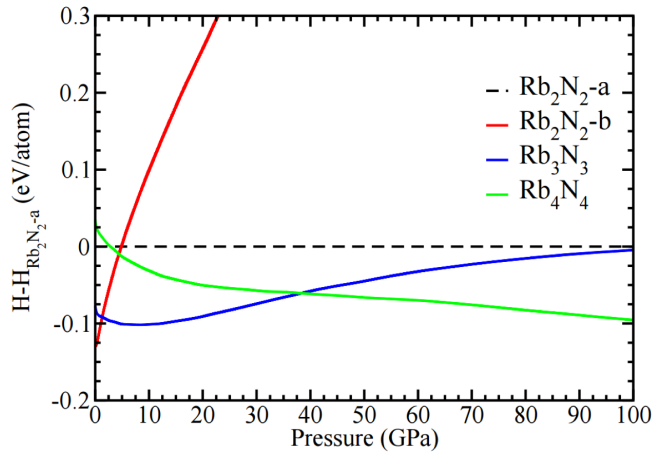

FIG. 3. Enthalpy difference between Rb$_3$N$_3$-*P-1* phase and other phases of crystals with RbN stoichiometry (Rb$_2$N$_2$-*P-1*-a from the 0 GPa search, Rb$_2$N$_2$-*P-1*-b phase transition found during the compression study, and Rb$_4$N$_4$-*P-1* from the 60 GPa search) showing where each phase is thermodynamically favorable.

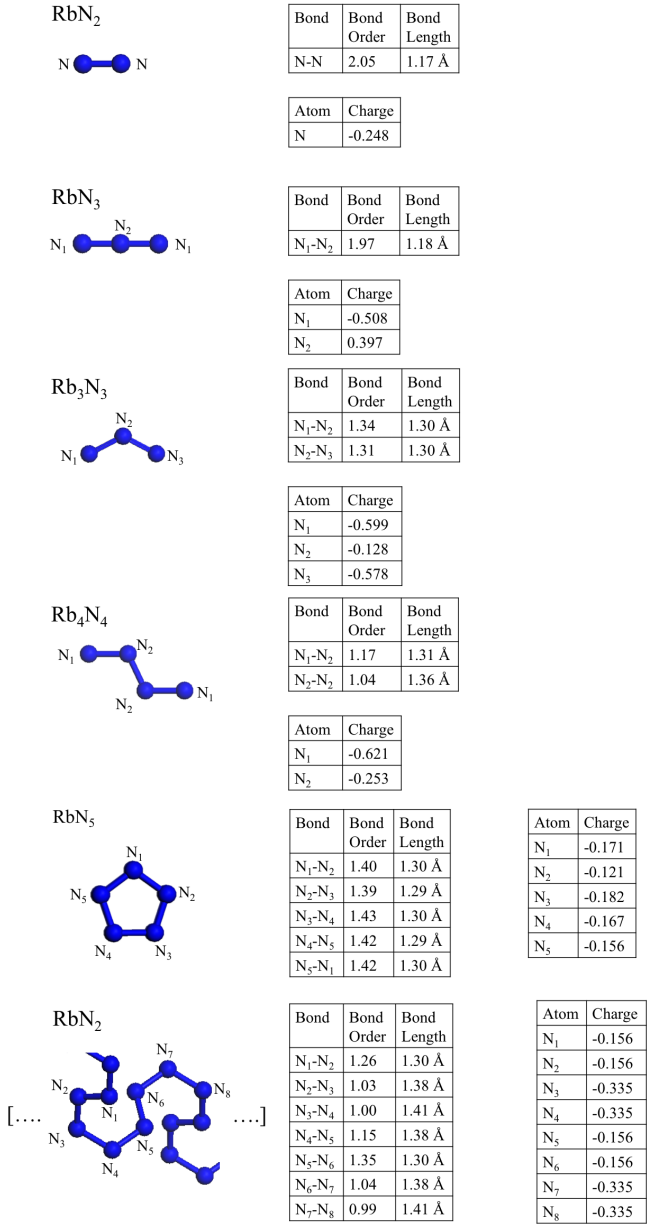

FIG. 4. Charges on the nitrogen atoms and the N-N bond orders for several $Rb_xN_y$ crystals calculated at 0 GPa.

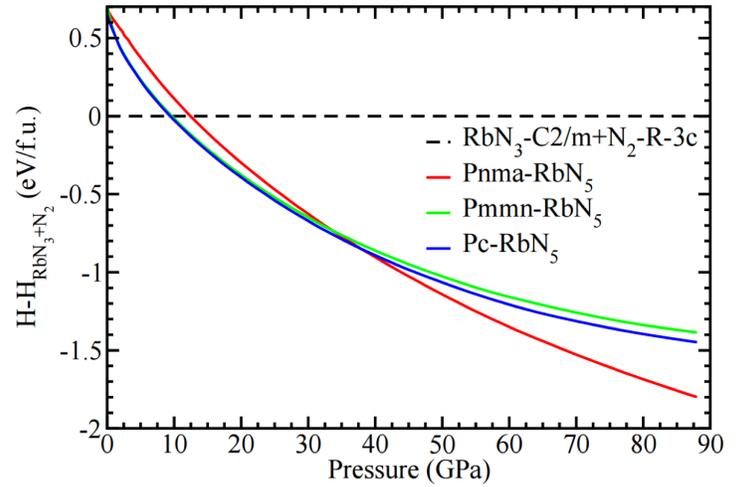

FIG. 5. Enthalpy difference between various phases of $RbN_5$ and the stoichiometric mixture of the precursors $RbN_3 + N_2$.

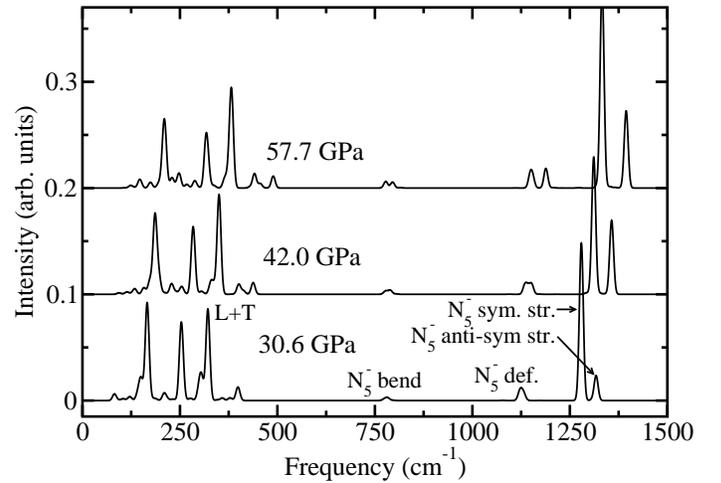

FIG. 6. The calculated Raman spectra of $RbN_5$-$Pnma$ from 30.6 GPa up to 57.7 GPa. Mode assignments are given at 30.6 GPa, which show lattice and librational modes (L + T), $N_5^-$ bending modes, $N_5^-$ deformation (def.) modes, and $N_5^-$ stretching (str.) modes.

The most nitrogen-rich compound, $RbN_5$ consisting of cyclo-$N_5$ pentazolates, can in principle be synthesized at high pressures by compressing rubidium azide ($RbN_3$) and nitrogen ($N_2$) mixture in a diamond anvil cell. Rubidium azide is well studied with several known phase transitions[40–42]. To investigate whether synthesis can be achieved at modest pressures, the enthalpy difference between the phases of $RbN_5$ and precursors $RbN_3$-$C2/m$ + $N_2$-$R$-$3c$ is calculated as a function of pressure and shown in Fig. 5. As $RbN_5$-$Pc$ is thermodynamically stabilized with respect to the precursors at 9.42 GPa, it can be synthesized upon compression above this pressure. Also seen from Fig. 5, $RbN_5$-$Pc$ phase undergoes a phase transition to $RbN_5$-$Pnma$ phase at 37.9 GPa, the latter being found in molecular search at 60 GPa.

As our work indicates a possibility of synthesis of $RbN_5$ upon compression of the mixture of rubidium azide $RbN_3$ and nitrogen $N_2$ precursors upon compression above 9.42 GPa, we calculated the pressure-dependent Raman spectra of $RbN_5$-$Pnma$ phase from 30.6 GPa up to 57.7 GPa to assist in experimental identification of pentazolates, see Fig. 6. $CsN_5$ was successfully synthesized upon compressing and laser heating the precursors to pressures near 60 GPa[12]. Due to the chemical similarities of Cs and Rb, it is likely that synthesis of RbN$_5$ might occur under similar conditions of pressures high enough to provide additional $PV$ enthalpy contribution to stimulate the chemical transformation. Due to this reason, the Raman spectrum of RbN$_5$-$Pnma$ phase, the lowest formation enthalpy pentazolate phase at higher pressures, is calculated and shown in Fig. 4. The low frequency region in the Raman spectrum at 30.6 GPa from 85 to 401 cm$^{-1}$ contains the lattice and N$_5^-$ librational modes (L + T). Although there are several other librational modes in this region, three of them stand out due to their strong intensity at 166 cm$^{-1}$, 253 cm$^{-1}$, and 322 cm$^{-1}$, see Fig. 6. In addition, there are several N$_5^-$ bending Raman modes with weak intensity at 781 cm$^{-1}$ at 30.6 GPa. However, these modes split at higher pressures into two clearly visible bands at 778 cm$^{-1}$ and 795 cm$^{-1}$ at 57.7 GPa. Another mode splitting is found at higher pressures for the N$_5$ deformational (def.) band at 1125 cm$^{-1}$ at 30.6 GPa, which splits into two clearly visible bands at 1151 cm$^{-1}$ and 1189 cm$^{-1}$ at 57.7 GPa. The two highest frequency bands are made of symmetry (sym.) and anti-symmetric breathing or stretching (str.) modes as seen in Fig. 6. All the calculated modes show normal pressure behavior with increasing pressure: the mode frequencies blue shift and their character does not change upon compression.

## IV. CONCLUSION

Three novel high-nitrogen content Rb$_x$N$_y$ compounds are discovered using first principles evolutionary structure search. The first, rubidium pentazolate RbN$_5$-$Pc$ consisting of cyclo-N$_5^-$ anions is predicted to be thermodynamically stable at pressures exceeding 15 GPa. Heating and compressing the mixture of rubidium azide RbN$_3$ and N$_2$ gas is expected to produce RbN$_5$ as our calculations demonstrate that RbN$_5$-$Pc$ becomes thermodynamically preferred compared to the stoichiometric mixture of the precursors at pressures above 9.42 GPa. Raman spectrum of rubidium pentazolate RbN$_5$-$Pnma$ is calculated to assist in experimental identification of N$_5^-$ modes, which should be observed in experiment if such compound appears upon its synthesis. The second high-nitrogen content compound, RbN$_2$-$Pbcn$, consisting of single bonded infinite nitrogen chains becomes thermodynamically stable at 60 GPa. The third newly discovered compound, Rb$_4$N$_6$-$P$-$1$, consists of N$_6$ hexazine rings is thermodynamically stable at 60 GPa. In addition, a compound Rb$_3$N$_3$-$P$-$1$ consisting of bent N$_3$ molecules, is predicted to be stable between 30 and 60 GPa. Although there are some similarities in structures of Rb polynitrogens with those found for other alkali polynitrides, the N$_3$ anions in Rb$_3$N$_3$-$P$-$1$ phase have not been seen before.

## ACKNOWLEDGMENTS

The research is supported by the Defense Threat Reduction Agency, (grant No. HDTRA1-12- 1-0023) and Army Research Laboratory through Cooperative Agreement W911NF-16-2-0022. Simulations were performed using the NSF XSEDE supercomputers (grant No. TG-MCA08X040), DOE BNL CFN computational user facility, and USF Research Computing Cluster supported by NSF (grant No. CHE-1531590).